%% file: 1_review.tex
\documentclass[review]{elsarticle}

\usepackage{hyperref}
\usepackage{xcolor}
\usepackage{ulem}
\usepackage{framed}
\usepackage[utf8]{inputenc}
\usepackage{graphicx}
\usepackage{pifont}

\definecolor{blue}{rgb}{0.74, 0.83, 0.9}
\definecolor{shadecolor}{named}{blue} 

\journal{Journal of \LaTeX\ Templates}

\input{6_Abbreviations}

\begin{document}

\begin{frontmatter}

\title{Towards rational glyco-engineering in CHO: from data to predictive models}

\author[boku,acib]{Jerneja {\v S}tor}
\author[acib,uvie]{David E. Ruckerbauer}
\author[acib,uvie]{Diana Sz{\'e}liov{\'a}}
\author[uvie,acib]{J\"urgen Zanghellini\corref{coraut}}
\ead{juergen.zanghellini@univie.ac.at}
\author[boku,acib]{Nicole Borth\corref{coraut}}
\ead{nicole.borth@boku.ac.at}

\cortext[coraut]{Corresponding authors}

\address[boku]{Department of Biotechnology, University of Natural Resources and Life Sciences Vienna, A-1190 Vienna, Austria, EU}
\address[acib]{acib - Austrian Centre of Industrial Biotechnology, Krenngasse 37/2, A-8010 Graz, Austria, EU}
\address[uvie]{Department of Analytical Chemistry, University of Vienna, A-1090 Vienna, Austria, EU} 

\begin{abstract}
Metabolic modeling strives to develop modeling approaches that are robust and highly predictive. To achieve this, various modeling designs, including hybrid models, and parameter estimation methods that define the type and number of parameters used in the model, are adapted. Accurate input data play an important role so that the selection of experimental methods that provide input data of the required precision with low measurement errors is crucial. For the biopharmaceutically relevant protein glycosylation, the most prominent available models are kinetic models which are able to capture the dynamic nature of protein N-glycosylation. In this review we focus on how to choose the most suitable model for a specific research question, as well as on parameters and considerations to take into account before planning relevant experiments.

\end{abstract}


\end{frontmatter}

\input{2_Introduction}
\input{3_Models}
\input{4_Methods}
\input{5_Conclusions_acknowledgements}

\bibliography{1_review}

\end{document}

%% file: 6_Abbreviations.tex
\usepackage[toc,section=section,acronym,nonumberlist,nogroupskip]{glossaries}
\usepackage{xspace}

\newacronym{cho}{CHO}{Chinese hamster ovary}
\newacronym{ns}{NS}{nucleotide sugar}
\newacronym{ms}{MS}{mass spectrometry}
\newacronym{lc}{LC}{liquid chromatography}
\newacronym[firstplural=monoclonal antibodies (mAbs)]{mab}{mAb}{monoclonal antibody}
\newacronym{ga}{GA}{Golgi apparatus}
\newacronym{er}{ER}{endoplasmatic reticulum}
\newacronym{gfa}{GFA}{Glycosylation Flux Analysis}
\newacronym{igg}{IgG}{imunoglobulin G}
\newacronym{fba}{FBA}{Flux Balance Analysis}

\newcommand{\cho}{\gls{cho}\xspace}
\newcommand{\ns}{\gls{ns}\xspace}

\newcommand{\Nss}{\Glspl{ns}\xspace}
\newcommand{\nss}{\glspl{ns}\xspace}
\newcommand{\ms}{\gls{ms}\xspace}
\newcommand{\Ms}{\Gls{ms}\xspace}
\newcommand{\lc}{\gls{lc}\xspace}

\newcommand{\ga}{\Gls{ga}\xspace}
\newcommand{\gfa}{\Gls{gfa}\xspace}
\newcommand{\igg}{\Gls{igg}\xspace}
\newcommand{\fba}{\Gls{fba}\xspace}
\newcommand{\er}{\Gls{er}\xspace}

%% file: 2_Introduction.tex
\section{Introduction}
High productivity and correct glycosylation are two main goals of the pharmaceutical industry for biotherapeutics production. Glycosylation is one of the most important quality parameters. Due to (i) its impact on biological activity, \textit{in vivo} half-life and immunogenicity, all of which define the safety and the efficacy of the product and (ii) the non-template driven complex network of the glycosylation machinery it needs to be carefully monitored and regulated \cite{hajba_glycosylation_2018, kontoravdi_computational_2018}. The most widely used expression systems for biotherapeutics are \cho cell lines \cite{walsh_biopharmaceutical_2018}. \cho cells are able to produce human-like glycosylation patterns, which make therapeutic proteins more compatible with and bioactive within the human body \cite{lalonde_therapeutic_2017}. Despite being commonly used, \cho cells still struggle with instability caused by mechanisms on genomic, transcriptomic and proteomic level \cite{dahodwala_fickle_2019} and high productivity often results in simplified glycosylation due to overload of the cellular glycosylation machinery \cite{sou_how_2015, jimenez_del_val_dynamics_2016, Bydlinski2020}.Moreover, non-engineered \cho cell lines are not able to produce some human glycosylation structures such as $\alpha$-2,6-sialylation \cite{Onitsuka2012} and $\alpha$-1,3/4-fucoslyation \cite{Howard1987}. Rarely, they can produce glycosylation structures not present in humans, such as N-glyconeuramic acid (Neu5Gc) \cite{Ghaderi2010} and galactose-$\alpha$1,3-galactose \cite{Bosques2010}, which can induce immune response in humans. Nevertheless, after careful selection of suitable subclones, \cho is still the most adequate expression system \cite{walsh_biopharmaceutical_2018}. In comparison to human cell lines, it is safer to use due to lower susceptibility to human viruses. Compared to other available mammalian cell lines, the \cho glycosylation machinery produces the most human-like glycosylation profiles, which is, as mentioned above, important for immuno-compatibility within the human body. Non-mammalian cell lines that lack the required glycosylation ability are thus mainly used for production of non-glycosylated products \cite{lalonde_therapeutic_2017}.\\
Glycosylation, the attachment of sugar moieties to proteins, is a posttranslational modification which takes place in the \er and the \ga. First, a pre-made complex attached to dolichol is transfered to the nascent polypetide. Subsequently, after removal of some sugar moieties, \er and \ga resident glycosyltransferases attach additional sugar residues from \ns to the growing glycans \cite{varki_chapter1_2015}. The pattern of glycosylation of proteins is influenced by the producer cell line \cite{lakshmanan_multiomics_2019, fan_amino_2015}, culture conditions \cite{st_amand_identifying_2016, naik_impact_2018} \cite{zhang_glycan_2020}$^*$ and the structure and amino-acid sequence of the protein \cite{suga_analysis_2018, chung_combinatorial_2017, losfeld_influence_2017}. For example, the therapeutically important antibodies, like IgG1, bear relatively simple biantennary glycans at a conserved N-glycosylation site at Asn297 in each Fc region \cite{wang_antibody_2018, sha_n-glycosylation_2016}, while other proteins, such as erythropoeitin have several glycosylation sites and also bear more complex structures with up to four antennae \cite{Sasaki1988}.\\
Glycoform heterogeneity at the time of harvest, called ``temporal dynamics'' is influenced by the kinetics of metabolic reactions involved in glycosylation which in turn is primarily determined by changes in availability of nucleotide sugar donors and enzyme co-factors, the levels of glycosylation enzymes and their activity and by-products from the central energy metabolism \cite{sumit_dissecting_2019}. As protein glycosylation is a complex, multi-step process taking place in several cellular compartments that the protein reaches sequentially, this non-template mechanism as such can lead to considerable variability. Consistent glycosylation between production batches is important to meet safety specifications of the product \cite{st_amand_identifying_2016}, as changes in glycosylation influence its pharmaceutical properties \cite{majewska_n-glycosylation_2020}$^*$.\\
Protein glycosylation can be manipulated with different glycoengineering approaches that include genetic engineering of cell lines \cite{stach_model-driven_2019}$^*$ \cite{krambeck_model-based_2017}, modifications of cell culture media \cite{zhang_glycan_2020,kotidis_model-based_2019} $^*$\cite{karst_modulation_2017, hutter_glycosylation_2017} and process parameters \cite{sou_model-based_2017} which is also reviewed in Sha et al. \cite{sha_n-glycosylation_2016}. Due to the complexity of the glycosylation machinery, the outcomes of these interventions do not always show the desired results. Yet it is important to understand the connecting mechanisms between intracellular factors associated with glycan biosynthesis pathways and the consequences of glycoengineering, media composition and process parameters. For this, further mechanistic insight into the cell is required to explore intracellular changes \cite{sha_n-glycosylation_2016}. Considering the complexity of the glycosylation process and many parameters that play a role, it is difficult to investigate the process just by experimental work and this is where mathematical models offer clear advantages and insight. They complement experimental investigations in addressing the effects of cellular or process changes induced to increase recombinant protein productivity or maximize desired glycoform fractions within the glycoform population \cite{tejwani_glycoengineering_2018}. \\
Here we are interested in the impact of process conditions on glycosylation. For this, we need to build a model that contains the glycosylation process coupled with a process model that allows us to understand the multiple interactions and feed-back loops between the two. Each of these "modules" can be modelled differently and has different data requirements. We will address three major topics, namely (i) an overview of the recent developments in glycosylation modelling and glyco-engineering with a focus on kinetic modelling, (ii) experimental methods to measure relevant parameters required for valid predictions, and (iii) available data sources for these parameters. The focus is to enable the choice of a model most suitable for a specific research question, but also the planning of relevant experiments.

%% file: 3_Models.tex
\section{Recent advances in glyco-engineering and -modelling}
The first glycosylation model was established in 1995 by Shelikoff et al. \cite{shelikoff_modelling_1996}. Since then and with the increase in therapeutic protein production, the interest in glycosylation modelling has grown in academia and industry \cite{zhang_glycan_2020, stach_model-driven_2019, kotidis_model-based_2019, kotidis_harnessing_2020}$^*$\cite{krambeck_model-based_2017, karst_modulation_2017, hutter_glycosylation_2017, sou_model-based_2017, sha_prediction_2019, kremkow_glyco-mapper_2018}. These models are listed chronologically in Figure 1, distinguished by the respective modeling approach, kinetic or stoichiometric. Several deficiencies still constrain the models' usability, as (i) mechanisms of glycosylation are not yet fully understood \cite{jimenez_del_val_dynamics_2016}, (ii) a large number of kinetic and transport parameters used for developing the models are not fully known and may well be species-specific (we do not know for lack of experimental and precise data) \cite{, hossler_systems_2007} and (iii) after simulations and sensitivity analyses, the predictions have to be compared to multiple experiments, to confirm their predictive power \cite{jimenez_del_val_dynamic_2011, krambeck_mathematical_2005}.

\begin{figure}[ht]
    \centering
    \includegraphics[width=12cm]{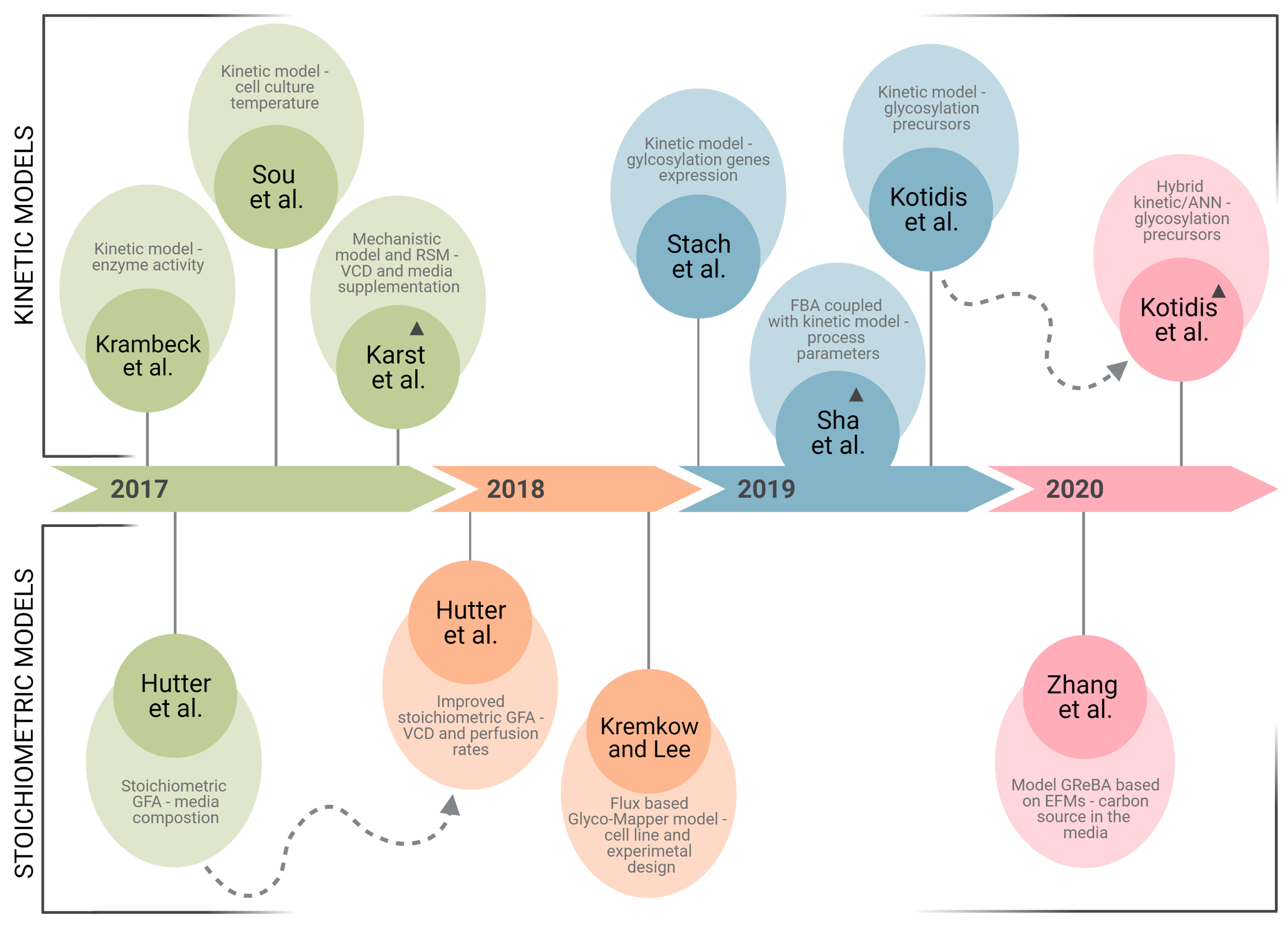}
    \caption{Reviewed models for predicting protein glycosylation in \cho in chronological order and separated by the respective modeling approach, kinetic or stoichiometric: in lighter colors [type of model]-[effect on glycosylation profile]; dashed arrow represents continuing work on the model; GFA - Glycosylation Flux Analysis; RSM - Response Surface Model; VCD - Viable cell density; FBA -  FLux Balance Analysis; ANN - Artificial Neural Network; GReBA - Glycan Residues Balance Analysis; EFM - Elementary Flux Mode; \ding{115} Publications include other modeling approaches besides kinetic, see chapter 2.2 for further details. Created with BioRender.com}
    \label{ModelsTimeline}
\end{figure}

\subsection{Modelling approaches}
Mathematical modelling can take different approaches depending on the system under investigation and the scope of the system. In general, mathematical modelling can be divided into stoichiometric and kinetic modelling, where each has its own strengths and weaknesses. In brief, stoichiometric modelling assumes steady state and it can be used on genome-scale, whereas kinetic modelling can describe dynamic changes, but  can only be performed on smaller scale, for example representing separate cellular processes. For a detailed description of the modelling approaches we refer the reader to recent review papers on metabolic modelling \cite{sha_mechanistic_2018, kyriakopoulos_kinetic_2018, galleguillos_what_2017}.

As glycosylation is a dynamic process that changes over time, the best way to describe it is with kinetic modelling. This type of model requires kinetic parameters which add up to the complexity, so that it is not suitable for large genome-scale models \cite{sha_mechanistic_2018}. To develop a kinetic model with good predictive capability, reliable and precise input data on essential elements are required, including (i) a defined model structure and reaction network, (ii) parameters for kinetic reactions, (iii) an estimation of unknown parameters and (iv) experimentally obtained mass balances \cite{kyriakopoulos_kinetic_2018}. A schematic representation of the process of building a model is shown in Figure 2. The first step is to define a model structure with a detailed cellular network and the metabolic pathway under investigation (Figure 2a). The defined model is then fed with input data from experimental work and kinetic parameters obtained from the literature (Figure 2b). The next step is to train the model using different approaches to adapt the quality of and the precise parameters of the input data (Figure 2c) which in the end give us accurate predictions (Figure 2d). The circle can be repeated to include new data, repeat training of the model and obtain predictions that more accurately represent the actual events in the cell.

\ga, as the center point of glycosylation, includes complex dynamics for which three different modelling concepts were proposed so far and were also mainly used in model designs. The first theory from Uma\~{n}a and Bailey proposed four well mixed reactors in series based on vesicular transport theory with static containers where protein flows through, but enzymes are static \cite{umana_mathematical_1997}. Later, Hossler et al. included \ga cisternal maturation theory and introduced four plug-flow reactors in series where \ga compartments were no longer static, but rather dynamic \cite{hossler_systems_2007}. The latest approach, proposed by Jimenez del Val et al., added the recycling of resident proteins in the \ga and developed a single plug flow reactor design \cite{jimenez_del_val_dynamic_2011}. All three types of models are frequently used in glycosylation models where \ga is modeled in more detail. As explained in Krambeck et al., the latter two give a better presentation of \ga. Nevertheless, all three are an idealized presentation of the \ga environment \cite{krambeck_model-based_2017}.

\begin{figure}[ht]
    \centering
    \includegraphics[width=12cm]{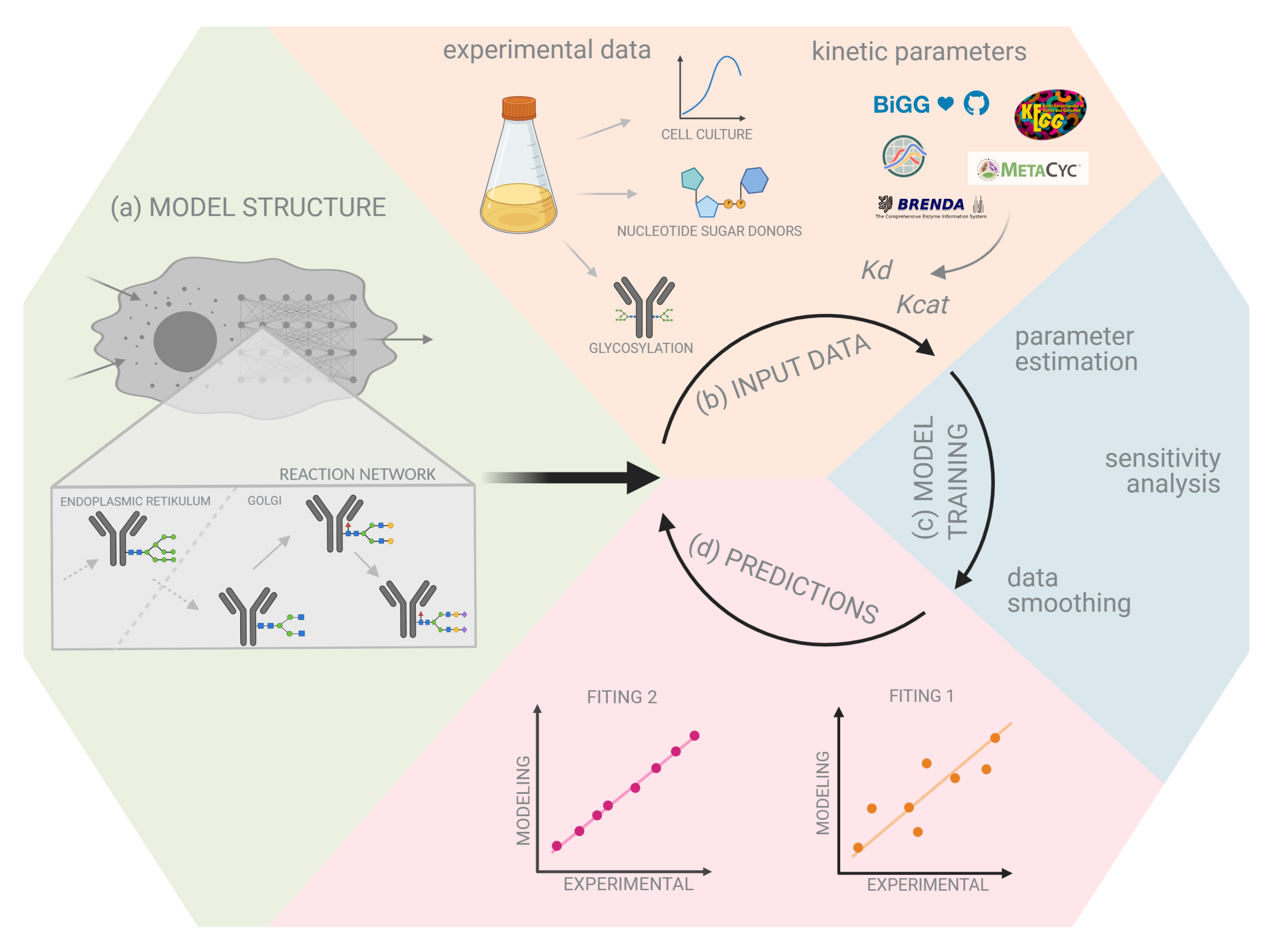}
    \caption{A workflow for establishing a metabolic model which starts with (a) definition of the model structure with detailed cellular network and the selected cell process reaction network. The flow continues to (b) feeding input data from experimental work or the literature and (c) different steps for training the model to provide accurate (d) predictions. Created with BioRender.com}
    \label{ModelWorkflow}
\end{figure}

\subsection{Modelling designs and accuracy of predictions}
The most important attribute of metabolic models is their predictive capability. Reviewed models create predictions on glycosylation profiles based on different variable inputs, for example culture temperature, media supplementation and composition or gene expression (Figure 1). However, across publications there is no uniform way of assessing prediction accuracy which would allow a good comparison of available models. Sha et al. combined stoichiometric modelling with kinetic, where they linked \fba with a kinetic model. The \fba served to provide \nss fluxes which were then inputs for the kinetic model. The model resulted in a deviation between the simulations and experimental measurements of less than 10\% \cite{sha_investigation_2019}$^*$. Zhang et al. developed a Glycan Residues Balance Analysis (GReBA) model based on Elementary Flux Modes (EFMs) which predicted changes in N-glycosylation based on different carbon source in the medium. Using the EFM approach allowed them to simplify the reaction network and kinetic parameters which gave satisfying results with low absolute errors \cite{zhang_glycan_2020}$^*$. The model from Kotidis et al. yielded a $\pm$5\% accuracy regarding the distribution of glycoforms depending on the changes in glycosylation precursors feed into the media \cite{kotidis_model-based_2019}$^*$. They further developed and updated the model to a machine learning Artificial Neural Network (ANN). This type of data-driven models do not require extensive knowledge of the biological process or construction of the reaction network and require only minimal parameter estimation. Their hybrid kinetic/ANN model was able to give better predictions for glycoform distributions compared to previous kinetic models \cite{kotidis_harnessing_2020}$^*$. Kremkow and Lee provided a good evaluation of their Glyco-Mapper model which has 96\% accuracy, 85\% sensitivity and 97\% specificity in predicting glycoform patterns in response to glycosylation manipulations. However, it does not predict glycan concentrations \cite{kremkow_glyco-mapper_2018}. Another model that did not focus on quantification of glycoforms was developed by Krambeck et al. The model predicts the effect of changes in glycosylation enzyme activities on glycoforms observed which can be used for designing glycoengineering strategies for CHO \cite{krambeck_model-based_2017}. Sou et al. established a model that predicts the effect of culture temperature on N-glycosylation. The predictions agreed well with experimental measurements, more precisely the simulated and measured data were within a 3.8\% range \cite{sou_model-based_2017}. The work from Karst et al. included two types of models, a mechanistic mathematical model and a Response Surface Model, both predicting N-linked glycosylation under the effect of varied cell density and media supplementation. As both models were able to give satisfying predictions of glycan fractions within experimental parameters, the mechanistic model was also able to reliably predict the glycoforms in the extrapolation range from the culture conditions \cite{karst_modulation_2017}. Stach et al. took a different approach by coupling kinetic modelling with synthetic biology where they worked in the direction of predicting the effect of over-expression of glycosylation genes on N-glycosylation. They were able to increase the level of galactosylation on secreted \igg, and the model also performed well at predicting over-expression results. However, the predictions were limited by the lack of experimental data on the level of over-expression and the kinetic effects of over-expression \cite{stach_model-driven_2019}$^*$. Hutter el al. developed a \gfa model which is, compared to the above kinetic models, a stoichiometric model. It predicted fluxes in the glycosylation network that were in good agreement with the measured fluxes. However, as the model assumes a pseudo steady state, it cannot describe dynamic changes and it cannot directly predict the effect of changes in process parameters or genetic modifications on local enzyme-specific changes and global alterations in cell metabolism \cite{hutter_glycosylation_2017}. They further expanded their \gfa model and improved its computational efficiency which allowed them to identify process parameters that could contribute to the changes in the intracellular \igg glycosylation network. However, the improved \gfa still could not solve the dynamic enzyme-specific changes \cite{hutter_glycosylation_2018}.

The selection of the most suitable model will depend on the predictions we want to make as well as on the level of the details required to describe the glycosylation process. The majority of glycosylation models developed in recent years took kinetic modelling approaches due to their ability to describe dynamic changes in the process. From the above examples one can see that kinetic modelling is not yet fully optimized to provide precise predictions, yet it is still the preferred way to describe the dynamics of glycosylation. As kinetic modelling brings in additional complexity, a promising way forward is to construct hybrid models and to precisely evaluate which parts of the process need to be described in a dynamic manner. 

\subsection{How to reduce variation originating from model design}
In the design of the model and its corresponding metabolic network, a number of parameters is included, however, this number does not necessarily correlate to model predictive power. The number of parameters can be reduced by parameter estimation and sensitivity analysis which evaluates parameters for their impact on the model prediction and thus can point to inessential parameters. Thus parameters with minor influence on the model can be excluded without altering the output of the model \cite{Kastelic2019}. Another approach to avoid overparameterization is to assume constant values for parameters throughout the culture, to not estimate intracellular concentrations of metabolites \cite{kotidis_model-based_2019}$^*$ or in the case of concentrations of \nss, to assume that their concentrations are equal in cytosol and \ga \cite{sou_model-based_2017}. The validity of such assumptions can only be estimated, however, without precise input data and measurements. Selecting appropriate parameter combinations by excluding the ones with lower accuracy can lower the variation of model predictions \cite{stach_model-driven_2019}$^*$. Moreover, smoothing data makes the model more robust in case of errors in experimental measurements or outliers and helps to reduce the effect of noise coming from the experimental data \cite{hutter_glycosylation_2017}. 

All of the mentioned steps should be included in model training to decide on crucial and precise input parameters and avoid too complex models.

%% file: 4_Methods.tex
\section{Building blocks of glycosylation models and things to consider}
The models described in the previous section require experimental data input to adequately represent cell line or process specific performance, which typically includes cell culture parameters, i.e. viable cell density and viability, the consumption and secretion rates of metabolites, productivity of the selected protein and the glycoprofile of the product \cite{zhang_glycan_2020}$^*$ \cite{hutter_glycosylation_2017, sha_prediction_2019, kremkow_glyco-mapper_2018}. Some more complex models also require the quantification of nucleotides and \nss as precursors for glycosylation \cite{kotidis_model-based_2019}$^*$ \cite{karst_modulation_2017, sou_model-based_2017} and/or gene expression of glycosylation enzymes and transporter proteins involved in glycosylation \cite{stach_model-driven_2019}$^*$ \cite{sou_model-based_2017}. In the following we look into the methods available to analyze these parameters.

\subsection{Measuring N-glycosylation profile}
There are three options for analyzing glycosylation profiles of recombinant proteins: (i) analyze the intact glycoprotein, (ii) digest the protein to obtain glycopeptides to be analyzed or (iii) cleave the glycans from the protein and analyze. The first two options preserve the glycosylation site which is important for proteins with multiple glycosylation sites where glycosylation pattern may be site specific. Intact glycoprotein analysis additionally offers short analysis times and minimal sample preparation, but might face problems with detection of minor glycoforms. It can also be limited by the size of the protein, which can be avoided by digesting proteins and analyzing separate glycopeptides. On the other hand, analysis of released glycans, although analytically less challenging, is only suitable for proteins with a single glycosylation site, such as \igg, where it can provide an in-depth analysis of the glycan structure \cite{majewska_n-glycosylation_2020} $^*$ \cite{zhang_challenges_2016}. 

Structural complexity of glycans presents a challenge in glycoanalytics and so far there is no universal technique for analysis of all types of glycans. The traditionally used method is separation with \lc (eg. HPLC) coupled with detection by fluorescence or ultraviolet (UV) absorption. Compared to HPLC, \Ms is with higher sensitivity, shorter analysis times and high throughput data, gaining on its importance and becoming the most commonly used method. Other noticeable methods are nuclear magnetic resonance (NMR) spectroscopy, sequential \ms \((MS^n)\) and \lc. Recently developed methods are ion mobility-MS (IM-MS) and cryogenic IR spectroscopy, which can provide more detailed structural information (see review from Mucha et al. \cite{Mucha2019}$^*$).
In the previous modelling publications, combinations of high-throughput methods like MALDI-TOF \cite{krambeck_model-based_2017, kremkow_glyco-mapper_2018}, CGE-LIF \cite{karst_modulation_2017}, UPLC \cite{zhang_glycan_2020,stach_model-driven_2019}$^*$ \cite{hutter_glycosylation_2017}, capillary electrophoresis (CE) \cite{kotidis_model-based_2019}$^*$ \cite{sou_model-based_2017} or LC-MS/MS \cite{sha_investigation_2019}$^*$ were used. To our knowledge, no reports on problems related to precision of the measurements have been published, although these methods still have their own drawbacks as some cannot determine site specificity of distinguished isomers. Considering this, it was suggested that a way forward would be to use a combination of LC and MS methods \cite{Mucha2019}$^*$. Therefore when deciding on the appropriate method for glycosylation measurements, one should evaluate what the required input parameter for the model is, based on the relevance of the specific glycostructure for the target protein.   

\subsection{Measuring concentrations of nucleotides and nucleotide sugar donors}
\Nss are involved in glycosylation as donors of sugar moieties which are attached at the glycosylation site \cite{varki_2015}. Whereas some of the models involve experimental measurement of \nss, e.g. the study from Zhang et al. \cite{zhang_glycan_2020}$^*$ did not include that data. The authors argued, based on previous research \cite{villacres_low_2015, wong_investigation_2010}, that it is still challenging to accurately determine the concentrations of \nss and nucleotides. Furthermore, they argued that including this kind of data makes model simulations more challenging due to the higher number of unknown parameters \cite{hutter_glycosylation_2017}. Another option can be the approach taken by Sha et al. where they used  \fba to provide \nss fluxes which were used as an input for the kinetic model to predict N-glycosylation \cite{sha_prediction_2019}. Despite the analytical challenge, many models contain \ns measurements \cite{stach_model-driven_2019,kotidis_model-based_2019,kotidis_harnessing_2020}$^*$ \cite{karst_modulation_2017, sou_model-based_2017, kremkow_glyco-mapper_2018}.


There are different extraction methods available for which the reviews by Naik et al. (2018) and Rejzek et al. (2017) offer good overviews and evaluation. Separation and analysis of extracted \nss and nucleotides is challenging due to their chemical instability and nature, as they are extremely hydrophilic, acidic and polar molecules. Moreover, low levels of analytes, matrix retention, co-elution and poor retention in chromatography also hamper analysis. Similar to glycoanalysis, the most common method for analysis of \nss and nucleotides is HPLC coupled with \ms (LC-MS) or tandem \ms (LC-MS/MS) which offers high selectivity and specificity. Despite being a powerful method and its ability to detect unusual \nss, \ms has its drawbacks. It is not compatible with all separation methods, it faces problems with distinguishing isomers, byproducts formed during derivatization, low volatility reagents and stable ions that can be formed during the separation step. Another type of detection method are UV based methods, which can overcome these obstacles, but can only provide rapid and accurate quantification of well characterized \nss \cite{naik_impact_2018, rejzek_profiling_2017, qin_quantification_2018}. Besides the aforementioned methods, it is also possible to measure \nss with capillary electrophoresis \cite{bucsella_nucleotide_2016}, IP-RP (ion-pair reverse phase) HPLC \cite{sha_high-resolution_2020}$^*$ and FACE (Flourophore Assisted Carbohydrate Electrophoresis) \cite{barnes_isolation_2016}. Although many methods have been described, the analysis of \Nss and nucleotides remains a challenging task that still is limited in its precision and reliability. \nss and nucleotide data were still used in some recent models\cite{kotidis_model-based_2019}$^*$ \cite{karst_modulation_2017, sou_model-based_2017} due to their ability to connect the cellular metabolism with glycosylation.

\subsection{Kinetic data}
Glycosylation models typically also include kinetic data extracted from the literature. This adds connectivity between obtained experimental data and it is important for better predictive capability of the models \cite{sha_mechanistic_2018}. The main challenge of building a kinetic model is the need to include a significant amount of unknown kinetic parameters that are mostly system specific, but undetermined, which creates a challenge for parameter estimations and model simulations. On the other hand, modelling with kinetic data might provide a mechanistic insight for the dynamic changes of enzyme depended factors \cite{hutter_glycosylation_2017, hutter_glycosylation_2018}. Kinetic models \cite{zhang_glycan_2020,stach_model-driven_2019,kotidis_model-based_2019,sha_investigation_2019}$^*$ \cite{krambeck_model-based_2017, karst_modulation_2017, sou_model-based_2017, kremkow_glyco-mapper_2018} mostly implemented data on the distribution of glycosylation enzymes in \ga, the dimensions of \ga (volume and length), \ga compartments residence times, enzyme kinetic constants, enzyme activity parameters, enzyme dissociation constants, enzymatic reaction conditions (with corresponding reactants and products), enzyme concentrations, \ns concentrations in \ga, protein flux in \ga, and transporter protein distribution parameters. 

Databases available to obtain data for kinetic and stoichiometric models are BiGG (genome-scale models), KEGG (genes, enzymes, reactions, pathways), MetaCyc (enzymes, pathways), BioModels (established models from literature) and BRENDA (kinetic parameters for enzymes) (Figure 2b) \cite{sha_mechanistic_2018}. However, only a small portion of parameters in the kinetic models is obtained via experimental work for a specific cell line or cultivation system, and most of this data were collected in older research work using different cell lines. Therefore, before using these parameters in a CHO relevant context, they should be re-estimated and adjusted for the metabolic profile of the selected cell line \cite{sha_mechanistic_2018}. Another flaw, even of experimentally obtained enzymatic parameters, is that the values are measured from \textit{in vitro} studies, which is not completely representative of the enzyme nature in \textit{in vivo} conditions \cite{Garcia-Contreras2012}. In the case of unknown parameters, proficient knowledge of parameter estimation technologies is required for which there are several approaches available \cite{kotidis_model-based_2019}$^*$ \cite{jimenez_del_val_dynamic_2011, del_val_theoretical_2016, Ulonska2018}.

\subsection{Impact of experimental data precision on modelling predictions and how to handle this} 
Experimental analyses unavoidably have some measurement error. Széliova et al., analyzed the effect of sampling frequency and measurement accuracy of extracellular metabolites on model predictions, showing that low sampling frequency and high measurement error result in very inaccurate calculations of exchange rates, which in turn lead to inaccurate model predictions \cite{szeliova_error_2020}$^*$. Although this was done on the example of biomass and flux predictions, we suppose that it would have the same or similar effect on glycosylation models and it should therefore be taken into consideration when planning the experiments to provide input data for glycosylation models. \\

Currently, it is difficult to conclude which data should be included as on one hand it is needed to precisely describe the cellular environment which requires an extensive number of parameters and is possibly based on imprecise data, which adds complexity to the model and does not necessary improve the precision of the model. On the other hand, building a simpler model with fewer parameters might fail to describe important connections of the process and therefore cannot reach the precision we were striving for. Therefore one should carefully evaluate the data required and the precision of the methods available, include only the essential parameters and, in the case of experimental measurements of glycosylation profile and \nss, counteract the inaccuracy of analysis methods by appropriate sampling schemes and statistical tools \cite{Mucha2019}$^*$ \cite{rejzek_profiling_2017, qin_quantification_2018} and suitable extraction steps \cite{majewska_n-glycosylation_2020}$^*$ \cite{naik_impact_2018, zhang_challenges_2016, dietmair_towards_2010,  del_val_optimized_2013}. One of the possible improvements would be to generate a database of kinetic parameters relevant for \cho cell lines that would also be able to describe enzyme characteristics in \textit{in vivo} conditions. We also encourage to search for innovations in the field which offer more reliable measurements and also to apply established approaches from other expression systems. 

%% file: 5_Conclusions_acknowledgements.tex
\section{Conclusions}
Metabolic models require detailed knowledge, a significant amount of precise input data and their predictions are not error-free. The effect of these drawbacks can be minimized by advances in modelling designs and amongst others approaches, by generating input data with minimum error. Recent publications in the field present numerous models that are able to predict modifications in N-glycosylation in response to genetic engineering of the glycosylation pathway, changes in media supplementation such as carbon source and glycosylation precursors and changes in culture conditions. These advances now enable possible applications to improve control of glycosylation during a process, to find possible targets for genetic engineering of the glycosylation pathway, and to study the activity of single glycosylation enzymes and enzyme cascades, and to optimize both process parameters and media. 
\section{Acknowledgments}
This work was supported by the COMET center “acib: Next Generation Bioproduction”, which is funded by BMK,BMDW, SFG, Standortagentur Tirol, Government of Lower Austria and Vienna Business Agency in the framework of COMET - Competence Centres for Excellent Technologies. The COMET-Funding Program is managed by the Austrian Research Promotion Agency FFG. Additional funding came from the BOKU DocSchool PhD programm Bioprocess Engineering.